\documentclass{ws-procs9x6}
\usepackage{verbatim}
\usepackage{amsmath}

\newcommand{\mbf}[1]{\ensuremath\mbox{\boldmath{$#1$}}}

\newcommand{\lbe}[1]{\begin{equation} \label{#1}}
\newcommand{\ee}{\end{equation}}

\def\aj{AJ}%
\def\araa{ARA\&A}%
\def\apj{ApJ}%
\def\apjl{ApJ}%
\def\apss{Ap\&SS}%
\def\aap{A\&A}%
\def\prl{Phys.~Rev.~Lett.}%
\begin{document}

\title{NUMERICAL SIMULATIONS OF MULTI-SCALE ASTROPHYSICAL PROBLEMS:\\
  THE EXAMPLE OF TYPE Ia SUPERNOVAE}

\author{F. K. R{\"O}PKE}

\address{Max-Planck-Institut f{\"u}r Astrophysik,\\
Karl-Schwarzschild-Str.~1, D-85741 Garching, Germany\\
$^*$E-mail: fritz@mpa-garching.mpg.de\\
www.mpa-garching.mpg.de/\~{}fritz}

\begin{abstract}
Vastly different time and length scales are a common problem in
numerical simulations of astrophysical phenomena. Here, we present an
approach to numerical modeling of such objects on the example of Type
Ia supernova simulations. The evolution towards the explosion proceeds
on much longer time scales than the explosion process itself. The
physical length scales relevant in the explosion process cover 11
orders of magnitude and turbulent effects dominate the physical
mechanism. Despite these challenges, three-dimensional simulations of
Type Ia supernova explosions have recently become possible and pave
the way to a better understanding of these important astrophysical objects. 
\end{abstract}


\bodymatter

\section{Introduction}

Astrophysics naturally features problems on large scales, which often
can be addressed with the methods of hydrodynamics. The number of
particles is huge and the interactions are in many cases (with the
important exception of gravity) short-ranged. This allows the
description of the systems in terms of thermodynamical variables. From
the formation of planets to the evolution of large-scale structure in
the Universe, hydrodynamical methods have been successfully applied to
astrophysical problems on various spatial scales. 

Astrophysical problems usually challenge numerical techniques and
computational resources due to their pronounced multi-scale
character. Physical processes take place on vastly different time
scales. Moreover, the range of spatial scales
involved is typically far beyond the capabilities of today's
supercomputers. Therefore, approximations and numerical modeling are
inevitable.

Here, we will discuss the a typical astrophysical scenario -- the
thermonuclear explosion of a white dwarf (WD) star which is believed to
give rise to a Type Ia supernova (SN~Ia) explosion.
A comprehensive treatment of this scenario would involve the
modeling of the formation of the progenitor system, its stellar
evolution, its approach to the explosive state, the ignition of the
explosion, the explosion stage itself, and the evolution of the
remnant. But we are far from dealing with problems of such
complexity. The different stages of the evolution of the system are
characterized by very different timescales and distinct physical
mechanisms. For instance, the stellar evolution of the progenitor
system may take more than a billion years while the actual
explosion takes place on a timescale of seconds. Therefore, different
methods are applied to address these stages. Stellar evolution is
usually treated in hydrostatic approaches, the evolution towards the
ignition of the thermonuclear explosion takes about a century and needs
special hydrodynamical approximations, while the explosion process is
modelled via a combination of hydrodynamics, turbulence modelling and
treatment of nuclear reactions.

\section{Astrophysical Model}

The favored astrophysical model of SNe~Ia is the
thermonuclear explosion of a WD
star composed of carbon and
oxygen\cite{hoyle1960a,hillebrandt2000a}. This end stage of 
stellar evolution for intermediate and low-mass stars is stabilized by
the pressure of a degenerate electron gas, because after nuclear burning
of hydrogen and helium it fails to trigger carbon and oxygen
burning. A WD is a compact object which would be eternally stable, cool
off, and disappear from observations. However, since many stars live in
binary systems, it is possible that it accrets material from its
companion. There exists a limiting mass for stability of a degenerate
object like a carbon/oxygen WD (the Chandrasekhar mass $\sim$$1.38 M_\odot$) beyond which it
becomes unstable to gravitational collapse. Approaching the
Chandrasekhar mass, the density in the core of the WD reaches values
that eventually trigger carbon fusion reactions.

This leads to about a century of convective burning. Finally, however,
a thermonuclear runaway occurs and gives rise to the formation of a
burning front, usually called a thermonuclear flame. This flame
propagates outward burning most of the material of the star and
leading to an explosion -- a process that occurs on timescales of
seconds. 

Hydrodynamics allows for two distinct modes of flame
propagation\cite{landaulifshitz6eng}. While in a subsonic deflagration the flame is mediated by
the thermal conduction of the degenerate electrons, a supersonic
detonation is driven by shock waves. Observational constraints rule
out a prompt detonation for SNe~Ia\cite{arnett1971a} and the flame must
therefore start out in the slow deflagration mode\cite{nomoto1976a}. The flame
propagation has to compete with the expansion of the star due to the
nuclear energy release. Once the dilution due to expansion has lowered
the fuel density below a certain threshold, no further burning is
possible. The energy release up to this point needs to be sufficient
to gravitationally unbind the WD star and to lead
to a powerful explosion. This is only possible if the propagation
velocity of the deflagration flame is accelerated far beyond the speed
of a simple planar flame (the so-called laminar burning speed).

It
turns out that this can be achieved by the interaction of the flame
with turbulence.
This turbulence is generic to the scenario. Burning from the center of
the star outward, the flame leaves light and hot nuclear ashes below
dense and cold fuel. This inverse density stratification in the
gravitational field of the WD is buoyancy
unstable. Consequently, burning bubbles form and float towards the
surface. Shear flows at the interfaces of these bubbles lead to the
generation of turbulent eddies.  By wrinkling the flame these increase
its surface and the net burning rate is enhanced. Thus, the
flame accelerates.
Whether this acceleration suffices to yield the strongest SNe Ia
observed, is currently debated\cite{reinecke2002d,gamezo2004a}. It has been hypothesized that a
transition of the flame propagation from subsonic deflagration to
supersonic detonation in later stages of the explosion may occur and
provide an the ultimate speed-up of the flame\cite{khokhlov1991a}.

\section{Challenges}

The astrophysical scenario of SNe~Ia described in the
previous section obviously poses great challenges numerical
modeling. Many of the problems found here are typical for a broad
range of astrophysical phenomena. The contrast between the time scales
of the actual explosion to that of the ignition process (let alone
the stellar evolution of the progenitor) is only part of the
scale-problem. The spatial scale ranges in the explosion as well as in the
pre-ignition phase are huge. 
Both processes are dominated by
turbulence effects with integral scales not much below the radius of
the star ($\sim$$2000\, \mathrm{km}$). A typical Reynolds number is as
high as $10^{14}$ and consequently the Kolmogorov scale is less than a
millimeter. 
Turbulence effects with Reynolds numbers far beyond anything occurring
on Earth are common in astrophysics. The scales of the objects and
typical velocities are huge but at the same time the viscosities of
astrophysical fluids are not extraordinarily high.
This indicates that neither a full temporal nor spatial resolution in
a single numerical approach is possible. Therefore the problems are
usually broken down into sub-problems which can be treated with specific
approximations and numerical techniques.
Moreover, astrophysical equations of state are often more complex than
those found under terrestrial conditions and are in some cases not even
well-known.

%The post-explosion evolution of the supernova is dominated by
%radiation transport phenomena like many other astrophysical
%systems. This bears its own challenges and is addressed in
%E.~M{\"u}ller's contribution to this volume. 

\section{Governing Equations}

Hydrodynamical problems in astrophysics can often be treated with the
Euler equations with gravity as external force which have to be
augmented by a description of
nuclear reactions and an appropriate astrophysical equation of
state. This set of equations is obtained when reaction
and diffusive transport phenomena are
neglected:
\begin{eqnarray}
\frac{\partial \rho}{\partial t} &=& - \mbf{\nabla} \cdot (\rho \mbf{v}),\\
\frac{\partial \mbf{v}}{\partial t} &=& - (\mbf{v \nabla}) \cdot 
\mbf{v} - \frac{\mbf{\nabla} p}{\rho} + \mbf{\nabla}\Phi,\\
\frac{\partial \rho e_\mathrm{tot}}{\partial t} &=& - \mbf{\nabla} \cdot
(\rho e_\mathrm{tot} \mbf{v}) - \mbf{\nabla} (p \mbf{v}) + \rho
\mbf{v} \cdot \mbf{\nabla}\Phi + \rho S,\\
\frac{\partial \rho X_i}{\partial t} &=& - \mbf{\nabla} \cdot
(\rho X_i \mbf{v}) + \mbf{r}_{X_i}\\
\mbf{r}_{X_i} &=& f(\rho, T, X_i),\\
p  &=& f_{\mathrm{EOS}}(\rho,e_\mathrm{int},X_i),\\
T  &=& f_{\mathrm{EOS}}(\rho,e_\mathrm{int},X_i),\\
S &=& f(\mbf{r}),\\
\Delta \Phi &=& 4 \pi G \rho.
\end{eqnarray}
Mass
density, velocity, pressure, total energy, internal energy, mass
fraction of species $i$, temperature, reaction rates, chemical source term, and
gravitational potential are denoted by $\rho$, $\mbf{v}$, $p$,
$e_\mathrm{tot}$, $e_\mathrm{int}$, $X_i$, $T$, $\mbf{r}$, $S$, and
$\Phi$, respectively. 
Index $i$ runs over $1 \ldots N$, where $N$ is the number of species contained in the reacting
mixture. 

The equation of state is indicated by $f_\mathrm{EOS}$. For
astrophysical objects matter may occur under extreme conditions and
the equation of state may differ significantly form that of
terrestrial matter. This is the case for our example as the equation
of state of the WD star is dominated by an arbitrarily
relativistic an degenerate electron gas. The nuclei form an ideal gas
of nuclei, and radiation and electron-positron pair
creation/annihilation contribute to the equation of state as well.

In many cases effects like heat conduction and diffusion play a major
role. In principle this holds for the example considered here,
too. On the smallest scales, the flame is mediated by the heat
conduction of the degenerate electrons. However, these scales cannot
be resolved in full-star simulations and the treatment of flame
propagation we will discuss below parametrizes it in a way such that
the above set of equations is sufficient to describe the astrophysical
scenario. 

\section{Modeling approaches}

Different methods for modeling the hydrodynamics make different
approximations and are suitable to certain sub-problems or
simplifications of the problems. First and pioneering approaches to simulate SNe~Ia,
for instance, assumed spherical symmetry\cite{nomoto1984a}. These neglect
instabilities and turbulence effects, which have to be parametrized in
such simulations, but they allow for efficient Lagrangian
discretization schemes. Consistency and independence of artificial
parameters can, however, only be reached in multi-dimensional
simulations. Eulerian discretizations are preferred here.

Depending on the time scales of the physical phenomena under
considerations, certain simplifications can be made to the
hydrodynamical equations. While in the explosion simulations usually the set of
equations spelled out above is applied, the pre-ignition convection and ignition
processes, as well as the propagation of deflagration flames on small
scales, are strongly subsonic phenomena. The magnitude of steps allowed in the
numerical simulations is set by the fastest motions that contribute to
the mechanism. Therefore, for subsonic phenomena it would be an
overkill (and in some cases it would also be numerically unstable) to
follow sound waves. Therefore these are filtered out in anelastic and
low-Mach number approaches applied specifically to the pre-ignition
and ignition phase of thermonuclear supernovae. Allowing for much
larger time steps than the sound crossing time over a computational
grid cell, such approaches facilitate the numerical study of phenomena
taking place on time scales of minutes and hours. These approaches are
described in detail elsewhere \cite{kuhlen2006a,bell2004c} and will focus on the traditional
implementation of hydrodynamical processes below.

The problem of spatial scales is approached in two strategies. While
off-line small-scale simulations of otherwise unresolved phenomena
test the assumptions of large-scale models\cite{roepke2003a,roepke2004a,roepke2004b,schmidt2005e,zingale2005a}, these in turn rely on
models for unresolved effects. 
The outstanding challenge in the latter is the description of turbulence
effects and a promising strategy for addressing these in astrophysical
simulations is the application of subgrid-scale turbulence
models. Since in SNe Ia the propagation of the deflagration flame is
dominated by turbulence effects, such models are applied here and an
example will be discussed below.

The nucleosynthesis in astrophysical events is a rich phenomenon which
can involve hundreds of isotopes and reactions between them. While it
is possible to run extended nuclear reaction networks concurrently
with one-dimensional astrophysical simulations, they are prohibitively
expensive in three-dimensional approaches. Therefore, such simulations
usually apply greatly simplified treatments of nuclear reactions in
order to approximate the energy release. In this way the dynamical
effects of nuclear burning can be treated without large
errors. However, in order to compare the results of astrophysical
simulations with observables such as spectra and light
curves\footnote{the temporal evolution of the luminosity of the
  event, usually restricted to a range of wavelengths set by an
  observational filter} (the
only way to validate astrophysical models), details of the chemical
structure of the object are to be known. One approach to this issue is
to advect a number of tracer particles with the hydrodynamical
simulation which record the evolution of the thermodynamical
conditions, and to feed this information into extended nuclear
reaction networks in a postprocessing step\cite{travaglio2004a,roepke2006b}.

\section{Numerical Methods}

There exists a large number of standard techniques for solving the
Euler equations in hydrodynamical simulations. In astrophysics, a widely
used finite-volume approach that discretizes the integral form of the
equations, is the piecewise parabolic method\cite{colella1984a,fryxell1989a} -- based on a
higher-order Godunov scheme. 

The selection of the geometry of the computational grid needs special
consideration. Although spherical coordinates seem best suited for
many astrophysical objects featuring an average spherical symmetry,
these are afflicted with coordinate singularities. Therefore,
currently there seems to be a trend towards Cartesian set-ups.

The challenge of incorporating phenomena that occur on scales
unresolved in simulations has to be addressed by modeling. For the
example of large-scale SN~Ia simulations this applies to
the propagation of the thermonuclear flame and to turbulence. As
described above, both are connected. For modeling turbulence on
unresolved scales, a Large Eddy Simulation (LES) ansatz is
chosen. Flow properties on resolved scales are used to determine
closure relations for a balance equation of the turbulent velocity $q$ at
the grid scale\cite{schmidt2006c}.
%\begin{eqnarray*}
%&&\frac{\mathrm{D}}{\mathrm{D} t} - \frac{1}{\rho} \cdot (\rho
%  {l}_\kappa q_\mathrm{sgs} \mbf{\nabla} q_\mathrm{sgs}) -
%  {l} |\mbf{\nabla} q_\mathrm{sgs} |^2 = \frac{1}{\sqrt{2}}
%  C_\mathrm{A} g_\mathrm{eff} + {l}_\nu |S|^2 - \frac{7}{30}
%  q_\mathrm{sgs} d - \frac{q_\mathrm{sgs}^2}{{l}_\epsilon}\\
%  &&\frac{\mathrm{D}}{\mathrm{D} t} = \frac{\partial}{\partial t}
%  + \mbf{v} \cdot \mbf{\nabla},
%\end{eqnarray*}
%with the rate-of-strain scalar $|S|$, the divergence of the (resolved)
%velocity field $d$. The third term on the left-hand side of the
%equation describes subgrid-scale turbulent transport, while the terms
%on the right hand side account for the Archimedian force on subgrid
%scales, the energy transport from resolved to subgrid scales, and the
%rate of viscous dissipation. $l_\kappa$,
%$l_\nu$, and $l_\epsilon$ denote characteristic length scales which can
%be determined from closures associated with the 

The structure profiles of a thermonuclear flame at high and
intermediate fuel densities extent typically over less than a
centimeter. These scales cannot be resolved in simulations capturing
the evolution of the entire WD star (radius 2000 km and
expanding). Therefore, in these simulations, the flame is treated as a
mathematical discontinuity separating the nuclear fuel from the
ashes. A numerical technique to represent the propagation of this
discontinuity is the level set
method\cite{osher1988a,reinecke1999a}. It associates the flame surface 
$\Gamma(t)$ 
with the zero level set of function $G$:
$$
\Gamma(t) := \{\mbf{r} | G(\mbf{r},t) = 0\}.
$$
For numerical convenience, we require $G$ to be a signed distance
function to the flame front, $|\mbf{\nabla} G| \equiv 1$
with $G<0$ in the fuel and $G>0$ in the ashes. The equation of motion
is then given by
$$
\frac{\partial G}{\partial t} = (\mbf{v}_\mathrm{u} \mbf{n} +
s_\mathrm{u}) |\mbf{\nabla} G|.
$$
Here, $\mbf{v}_\mathrm{u}$ is the fluid velocity ahead of the flame,
$s_\mathrm{u}$ is the effective flame propagation velocity with
respect to the fuel, and $\mbf{n} = - \mbf{\nabla}G/|\mbf{\nabla} G|$
is the normal to the flame front.
This equation ensures that the zero level set (i.e.\ the flame) moves in normal
direction to the flame surface due to burning and additionally the
flame is advected with the fluid flow. The burning speed $s_\mathrm{u}$ has to be
provided externally in this approach since the burning microphysics is
not resolved. While the flame propagation proceeds with the well-known
laminar flame speed in the very first stages of the explosion, it
quickly gets accelerated by interaction with turbulence. By virtue of
the implemented subgrid-scale model the turbulent burning speed of the
flame can easily be determined. In the turbulent combustion regime
that holds in most parts of the supernova explosion, it is directly
proportional to the turbulent velocity fluctuations. This is the way
in which the multidimensional LES approach to flame propagation in
thermonuclear supernovae avoids tunable parameters in the description
of flame propagation.

In order to take the expansion of the WD into account in the
simulation, one has to adapt the computational grid accordingly. One
option is adaptive mesh refinement, which, however, suffers from the
usually volume-filling turbulent flame structure. Therefore, a
refinement all over the domain would be necessary and the gain in
efficiency from this method is marginal. An alternative is to use a
computational grid with variable cell sizes. This grid can be
constructed to track the expansion of the star\cite{roepke2005c} or the propagation of
the flame inside it (or both\cite{roepke2006a}).

\section{Three-dimensional Type Ia supernova simulations}

Applying the techniques discussed above, three-dimensional simulations of
deflagration thermonuclear burning can be
performed\cite{reinecke2002d,roepke2005b,schmidt2006c} (the
incorporation 
of a delayed detonation stage is also possible with slight
modifications of the methods\cite{roepke2007b}). The goals of such simulations is to
determine whether an explosion of the WD can be achieved in
the model and whether the characteristics of such an explosion meet
observational constraints. As a direct link from the simulation of the
pre-ignition convection studies and the flame ignition simulations to
the explosion models is still lacking, the flame ignition is
introduced by hand in configurations motivated by off-line studies\cite{garcia1995a,kuhlen2006a,iapichino2006a}. 

\begin{figure}
\begin{center}
\psfig{file=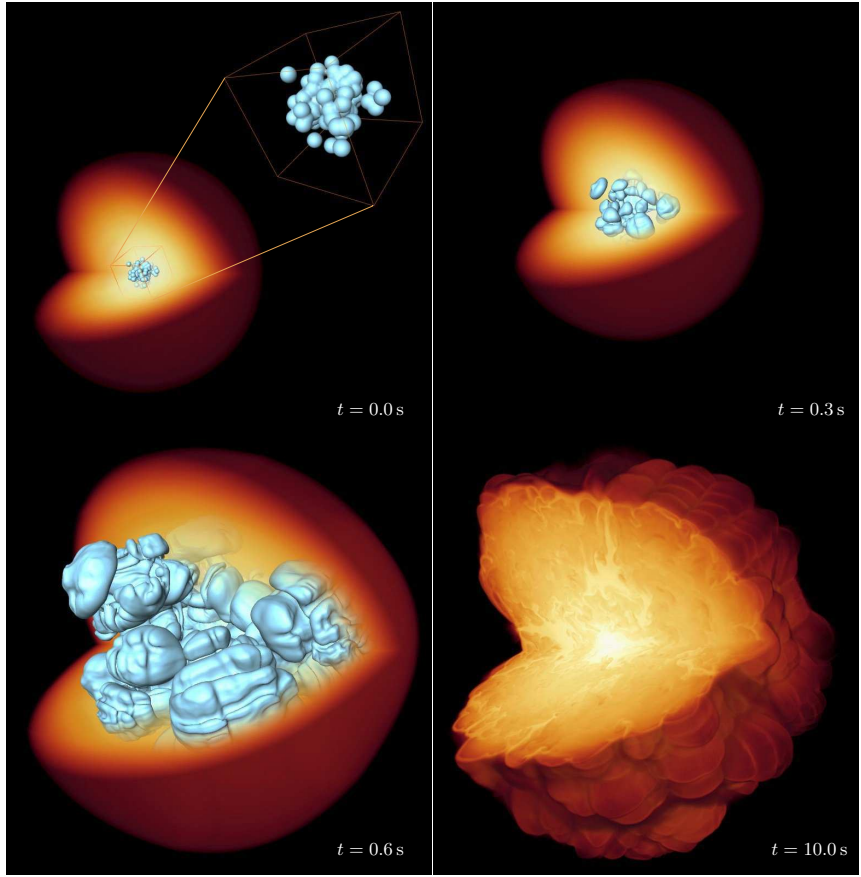,width=\textwidth}
\end{center}
\caption{Snapshots from a full-star SN~Ia simulation starting from
  a multi-spot ignition scenario. The density is
  volume rendered indicating the extend of the WD star and the
  isosurface corresponds to the thermonuclear flame. The last snapshot
  corresponds to the end of the simulation and is not on scale with
  the earlier snapshots.\label{fig:evo}}
\end{figure}

To illustrate the typical flame evolution in deflagration SN~Ia models,
the full-star model presented by
\cite{roepke2005b} shall be described here. The flame was ignited in a
number of randomly distributing spherical flame kernels around the
center of the
WD. This resulted in a foamy structure
slightly misaligned with the center of the WD (shown in Fig.~\ref{fig:evo}).
 Such multi-spot
ignition models are motivated by the strongly turbulent convective
carbon burning phase preceding the ignition (but alternatives such as
asymmetric off-center ignitions have also been considered).
Starting from this
initial flame configuration, the
evolution of the flame front in the explosion process is illustrated
by snapshots of the $G=0$ isosurface at $t = 0.3 \, \mathrm{s}$ and $t =
0.6 \, \mathrm{s}$ in
Fig.~\ref{fig:evo}. 
The development of
the flame shape from ignition to $t = 0.3 \, \mathrm{s}$ is
characterized by the formation of the well-known ``mushroom-like'' structures resulting
from buoyancy. This is especially well visible for the bubbles that
were detached from the bulk of the initial flame. But also the
perturbed parts of the flame closer to the center develop
nonlinear Rayleigh-Taylor features. 
During the following flame evolution, inner structures of smaller
scales catch up with the outer ``mushrooms'' and the initially separated structures
merge forming a more closed configuration (see snapshot at $t = 0.6 \,
\mathrm{s}$ of Fig.~\ref{fig:evo}). This is a result of the
large-scale flame advection in the turbulent flow, burning, and the expansion of
the ashes.
After about $2 \, \mathrm{s}$ self-propagation
of the flame due to burning has terminated in the model.
The subsequent evolution is characterized by the approach to
homologous (self-similar) expansion. The resulting density structure at the end of
the simulation is shown in the $t=10\, \mathrm{s}$ snapshot of
Fig.~\ref{fig:evo}. 

The goal of such simulations is to construct a valid model for SNe~Ia
which meets the constraints from nearby well-observed objects. Such
models can then be used to test and refine the methods that are used
to calibrate cosmological distance measurements based on SN~Ia
observations\cite{phillips1993a}, which pioneered the new cosmological standard model with
an accelerated expansion of the Universe\cite{riess1998a,perlmutter1999a} pointing to a dominant new
``dark'' energy form.

%\bibliographystyle{ws-procs9x6}
%\bibliography{astrofritz}

\end{document}